\documentclass[12pt,twoside]{article}
\usepackage{epsfig}
\usepackage{graphicx}
\usepackage{amssymb}
\usepackage{url}
\usepackage{amsfonts,amssymb,amsbsy}
\usepackage{psfrag}
\usepackage{amsmath}
\usepackage{comment}
\title{Skewon-Axion Medium and\\ Soft-and-Hard/DB Boundary Conditions}
\author{I.V. Lindell and A.H. Sihvola} 
\date{Department of Radio Science and Engineering\\ Aalto University, School of Electrical Engineering\\ P.O.Box 13000, Espoo 00076AALTO, Finland\\ {\tt ismo.lindell@aalto.fi, ari.sihvola@aalto.fi}}
\def\e{\begin{equation}} 
\def\f{\end{equation}} 
\def\ea{\begin{eqnarray}} 
\def\fa{\end{eqnarray}} 

\def\##1{{\bf #1\mit}}
\def\%#1{{\mbox{\boldmath $#1$}}}
\def\=#1{{\overline{\overline{\mathsf #1}}}}

\def\*{^{\displaystyle*}}

\def\.{\cdot}
\def\x{\times}
\def\oo{\infty}

\def\d{\partial}

\def\ra{\rightarrow}

\def\l#1{\label{eq:#1}}
\def\r#1{(\ref{eq:#1})}
\def\am{\left(\begin{array}{c}}
\def\amm{\left(\begin{array}{cc}}
\def\a{\end{array}\right)}

\def\A{\alpha}
\def\B{\beta}
\def\de{\delta}
\def\De{\Delta}
\def\E{\epsilon}
\def\g{\gamma}

\def\la{\lambda}

\def\M{\mu}
\def\o{\omega}

\def\t{\tau}
\def\z{\zeta}

\def\TH{\theta}

\def\VF{\varphi}

\def\ve{\%\varepsilon}

\def\tr{{\rm tr }}
\def\det{{\rm det}}

\def\W{\wedge}
\def\WW{\displaystyle{{}^\wedge}\llap{${}_\wedge$}}
\def\J{\rfloor}
\def\L{\lfloor}

\begin{document}

\maketitle

\begin{abstract}
The class of skewon-axion media can be defined in a simple and natural manner applying four-dimensional differential-form representation of electromagnetic fields and media. It has been recently shown that an interface of a uniaxial skewon-axion medium acts as a DB boundary requiring vanishing normal components of the D and B vectors. In the present paper a more general skewon-axion medium is considered. It is shown that a planar interface of such a medium acts as a boundary generalizing both soft-and-hard (SH) and DB boundary conditions to SHDB conditions. Reflection of a plane wave from a planar SHDB boundary is studied. It is shown that for the two eigenpolarizations the boundary can be replaced by equivalent PEC or PMC boundaries. The theory is tested with a numerical example. 
\end{abstract}

\section{Introduction}

The most general electromagnetic medium can be defined in terms of 36 parameters, either in terms of four medium dyadics as 
\e \am \#D_g\\ \#B_g\a = \amm \=\E & \=\xi\\ \=\z & \=\M\a \.\am \#E_g\\ \#H_g\a, \l{DB} \f
in three-dimensional representation for Gibbsian vector fields denoted by $\#D_g$ , $\#B_g$,$\#E_g$ and $\#H_g$, or in terms of a single medium dyadic as
\e \%\Psi = \=M|\%\Phi \l{PsiMPhi}, \f
applying four-dimensional differential-form representation for electromagnetic fields and media \cite{Deschamps,Difform}. In the latter case the field two-forms $\%\Psi$ and $\%\Phi$ can be expressed as
\e \%\Psi = \#D - \#H\W\#d\t,\ \ \ \ \%\Phi=\#B + \#E\W\#d\t, \l{PsiPhi}\f
in terms of three-dimensional (spatial) two-forms $\#D,\#B$ and one-forms $\#H,\#E$ when $\#d=\sum \ve_i\d_{x_i},\ i=1..4$ is the differential operator and $\t=ct$ is the normalized time. The medium dyadic $\=M$ mapping two-forms to two-forms corresponds to a $6\x6$ matrix in any basis expansion of two-forms. The notation in this paper follows that given in \cite{Difform}.
  
The medium dyadic $\=M$ has a natural (independent of any basis system) decomposition in three parts as
\e \=M = \=M_1 + \=M_2 + \=M_3 \f
and the dyadics $\=M_i$ are respectively called principal, skewon and axion components of $\=M$ \cite{Hehl}. The axion part is a multiple of the unit dyadic which in the notation of \cite{Difform} can be expressed as
\e \=M_3 = M_3\=I{}^{(2)T}, \l{M3}\f
while both $\=M_1$ and $\=M_2$ are trace-free dyadics. The components $\=M_1$ and $\=M_2$ can be defined so that the dyadics contracted by a quadrivector $\#e_N=\#e_{1234}$ as $\#e_N\L\=M_1$ and $\#e_N\L\=M_2$ are respectively symmetric and antisymmetric. Physical properties of these components are discussed in \cite{Hehl}.

The total number of 36 parameters is distributed by the three components so that the principal part $\=M_1$, corresponding to a trace-free symmetric $6\x6$ matrix has 20 parameters, the skewon part $\=M_2$, corresponding to an antisymmetric $6\x6$ matrix, has 15 parameters, and the axion part $\=M_3$, has 1 parameter. A medium consisting only of its axion parameter, $\=M=\=M_3$ in \r{M3}, has been called PEMC, perfect electromagnetic conductor, because it is a generalization of both PMC ($M_3=0$) and PEC ($1/M_3=0$), \cite{PEMC,PEMC1,AnnPhys}. The medium equations for the PEMC can be expressed in the simple form
\e \#D=M\#B,\ \ \ \ \#H=-M\#E. \l{PEMC} \f

\section{Skewon-Axion Medium}

One can show that the most general medium with no principal component, the skewon-axion medium \cite{Hehl02,Obukhov04,Hehl05} (also called as the IB-medium \cite{IB}), can be expressed in terms of a dyadic $\=B$, corresponding to a $4\x4$ matrix involving only 16 parameters, as
\e \=M = (\=I\WW\=B)^T= (\=B\WW\=I)^T. \l{MIB}\f
It will turn out that a plane wave in the most general skewon-axion medium is not governed by a dispersion equation, normally restricting the choice of the $\#k$ vector in a given medium. Thus, basically, one can choose any $\#k$ vector for a plane wave in such a medium. 

In \cite{IBDB} the problem of plane wave reflection from the planar interface of a uniaxial skewon-axion medium, defined by six medium parameters, was analyzed. It was shown that, despite the absence of any dispersion equation, the $\#k$ vector in the skewon-axion medium was uniquely determined by the interface conditions. Moreover, it was shown that the interface conditions serve as boundary conditions 
\e \#n\.\#D_g=0,\ \ \ \ \#n\.\#B_g=0, \l{DBcond}\f
for Gibbsian fields at the interface of the uniaxial skewon-axion medium, dubbed as the DB conditions \cite{DB,AP10}. It subsequently turned out that such conditions were defined already in 1959 \cite{Rumsey} and that they are known to have unique mathematical properties \cite{Yee,Kress}. The DB conditions have been shown to play a central role in electromagnetic cloaking problems \cite{Kong08,Yaghjian,Weder,Kildal09}. More recently, the same DB conditions were shown to emerge at the interface of a simple skewon medium defined by just one parameter in the medium conditions \cite{SS}
\e \#D=N\#B,\ \ \ \ \#H=N\#E, \l{N}\f
which differ slightly those of the PEMC (axion medium) in \r{PEMC}.

Fields in the skewon-axion medium can be compacly handled applying the four-dimensional formalism \cite{Difform}. The Maxwell equations outside sources are
\e \#d\W\%\Phi=0,\ \ \ \ \ \#d\W\%\Psi=0, \l{Max}\f
whence the field two-form $\%\Phi$ can be (locally) represented in terms of a potential one-form $\%\phi$ as
\e \%\Phi= \#d\W\%\phi. \f
Substituting \r{PsiMPhi} and \r{MIB}, the second Maxwell equation can now be expanded as 
\ea \#d\W\%\Psi &=& \#d\W(\=B{}^T\WW\=I{}^T)|(\#d\W\%\phi) \\
                &=& \#d\W(\=B{}^T|\#d)\W\%\phi) + \#d\W\#d\W(\=B{}^T|\%\phi) \\
                 &=& \#d\W((\=B{}^T|\#d)\W\%\phi) \\ 
                 &=& 0.\fa
This shows us that the general solution can be represented as
\e \%\phi = \#d\psi + \=B{}^T|\#d\VF, \f
in terms of two scalar functions $\psi$ and $\VF$, whence the field two-forms have the form
\ea \%\Phi &=& \#d\W\%\phi = \#d\W(\=B{}^T|\#d)\VF, \l{Phi1}\\
 \%\Psi &=& (\=B{}^T|\#d)\W\%\phi = \#d\W(\=B{}^{2T}|\#d)\VF. \l{Psi1}\fa

Instead of studying the general case, let us assume that the skewo-axion medium is defined by the dyadic
\e \=B = B\=I + \#a\ve_3 + \#b(\ve_1+ A\ve_4), \l{B}\f
where $\ve_1,\ve_2,\ve_3$ are spatial one-forms and $\ve_4=\#d\t$ a temporal one-form, and together they form a basis. The corresponding reciprocal basis vectors are denoted by $\#e_i$ with the property $\#e_i|\ve_j=\de_{ij}$ \cite{Difform}. 

Substituting \r{B} in \r{Phi1} and \r{Psi1}, the field two-forms can be shown to satisfy the conditions
\ea \ve_3\W(\ve_1+A\ve_4)\W\%\Phi &=&0, \\ 
    \ve_3\W(\ve_1+A\ve_4)\W\%\Psi &=&0, \fa 
for the special skewon-axion medium considered here. Inserting the expansions \r{PsiPhi}, the conditions become
\ea A\ve_3\W\ve_4\W\#B + \ve_3\W\ve_1\W\#E\W\ve_4&=&0, \\ 
    A\ve_3\W\ve_4\W\#D - \ve_3\W\ve_1\W\#H\W\ve_4&=&0, \fa 
and they are equivalent with the spatial conditions
\ea A\ve_3\W\#B + \ve_3\W\ve_1\W\#E&=&0, \l{3B}\\ 
    A\ve_3\W\#D - \ve_3\W\ve_1\W\#H&=&0. \l{3D}\fa 
To conclude, any fields in a skewon-axion medium defined by \r{B} and \r{MIB} must satisfy the conditions \r{3B} and \r{3D}.

\section{SHDB Boundary Conditions}

Let us interprete the result \r{3B} and \r{3D} in terms of three-dimensional Gibbsian vectors applying the rules given in the Appendix. Denoting again the Gibbsian vectors by the subscript $()_g$  we can write
\ea \#e_{123}|(\ve_3\W\#B) &=& \ve_3|(\#e_{123}\L\#B) = \#e_3\.\#B_g, \\
    \#e_{123}|(\ve_3\W\#D) &=& \ve_3|(\#e_{123}\L\#D) = \#e_3\.\#D_g, \\
 \#e_{123}|(\ve_3\W\ve_1\W\#E) &=&  \#e_3\.(\#e_1\x\#E_g) = \#e_2\.\#E_g, \\
 \#e_{123}|(\ve_3\W\ve_1\W\#H) &=&  \#e_3\.(\#e_1\x\#H_g) = \#e_2\.\#H_g. \fa
Here we have applied the property $\#e_i|\ve_j=\de_{ij}$ transformed to $\#e_i\.\#e_j=\de_{ij}$ for the spatial basis vectors $i,j=1,2,3$. 

Since the coefficient $A$ in the conditions \r{3B} and \r{3D} has the dimension of velocity (due to the normalized time parameter $\t=ct$), let us further express it as
\e A = Tc = \frac{T}{\sqrt{\M_o\E_o}}=\frac{\o T}{k_o} , \l{AT}\f
in terms of a dimensionless scalar $T$. Here we have tacitly assumed that the fields are time harmonic as $\exp(j\o t)$ with $k_o=\o\sqrt{\M_o\E_o}$. Thus, the conditions \r{3B} and \r{3D} can be represented in the following form for the Gibbsian field vectors,
\ea T\o\#e_3\.\#B_g + k_o\#e_2\.\#E_g&=&0, \l{3Bg}\\ 
    T\o\#e_3\.\#D_g - k_o\#e_2\.\#H_g&=&0, \l{3Dg}\fa 
and they are satisfied by any fields in the skewon-axion medium under consideration. One may note that the pair of conditions is invariant in the duality transformation $\#D_g\leftrightarrow\#B_g$, $\#E_g\leftrightarrow -\#H_g$ which leave the Maxwell equations invariant. 

Assuming now a planar interface at $\#e_3\.\#r=0$, with normal unit vector $\#e_3$, between the skewon-axion medium and an isotropic medium (parameters $\M_o,\E_o$), the conditions \r{3B} and \r{3D}, and their Gibbsian counterparts \r{3Bg} and \r{3Dg}, are actually continuous across the interface. Since these conditions at the interface are enough to determine the fields in the isotropic medium without having to solve for the fields behind the interface, they can be considered as boundary conditions. Although \r{3Bg} and \r{3Dg} were introduced through a consideration of a planar interface, their form is applicable to curved boundaries as well, with $\#e_3$ and $\#e_2$ denoting respective unit vectors normal and tangential to the boundary surface. 

The boundary defined by \r{3Bg}, \r{3Dg} appears as a generalization of both the DB boundary defined by the conditions \r{DBcond} and the soft-and-hard (SH) boundary \cite{SHS1,SHS2,Kildal09} defined by the conditions
\e \#e_2\.\#E_g=0,\ \ \ \ \ \#e_2\.\#H_g=0. \f
In fact, the SH conditions are obtained for the limiting parameter value $T\ra0$ while the DB conditions are obtained from \r{3Bg}, \r{3Dg} for $T\ra\oo$. For convenience, the more general boundary defined by \r{3Bg} and \r{3Dg} and combining both SH and DB conditions will be called the SHDB boundary. 

One can immediately observe that, due to its self-dual property, an object with rotational symmetry and SHDB boundary should appear invisible to the radar because the back-scattering cross section of such an object is zero \cite{AP09}.

\section{Gibbsian Medium Conditions}

The skewon-axion medium was defined in terms of four-dimensional quantities through \r{MIB}, \r{B}. For the convenience of readers not familiar with the four-dimensional formalism and for double-checking the results, let us now find the relation between the skewon medium interface and the boundary consitions \r{3Bg}, \r{3Dg} in terms of three-dimensional Gibbsian vectors and dyadics. For that purpose we first find the Gibbsian medium conditions by expanding
\e (\=B\WW\=I)^T= 2B\=I{}^{(2)T} - (\ve_3\W\=I{}^T\W\#a+ (\ve_1+ A\ve_4)\W\=I{}^T\W\#b), \f
and expand
$$ (\ve_3\W\=I{}^T\W\#a)|\%\Phi = (\ve_3\W\=I{}^T\W\#a)|(\#B+\#E\W\ve_4) $$
\e = \ve_3\W(\#a\J\#B) + (\ve_3\W\#E)(\#a|\ve_4)-(\ve_3\W\ve_4)(\#a|\#E). \f
The expansion for the second term is obtained similarly, replacing $\ve_3$ by $\ve_1+ A\ve_4$ and $\#a$ by $\#b$. Let us also expand the two vectors in their spatial and temporal components as 
\e \#a=\#a_s + a_4\#e_4,\ \ \ \ \#b=\#b_s+ b_4\#e_4, \f
With these we can split the medium equation
\e \#D-\#H\W\ve_4 = (\=B\WW\=I)^T|(\#B+ \#E\W\ve_4) \f
in its spatial and temporal parts and identify the fields as
\ea \#D &=& 2B\#B -\ve_3\W(\#a_s\J\#B)-\ve_1\W(\#b_s\J\#B) \nonumber \\ 
&& -(a_4\ve_3+b_4\ve_1)\W\#E, \\
 \#H &=& -A\#b_s\J\#B -(Ab_4+2B)\#E \nonumber\\
 &&-(\ve_3\#a_s+\ve_1\#b_s)|\#E.\fa

Applying the transformation rules given in the Appendix, the medium conditions can be written for the Gibbsian vectors as
\ea \#D_g &=& 2B\#B_g -\#e_3\x(\#a_s\x\#B_g)-\#e_1\x(\#b_s\x\#B_g)  \nonumber\\ 
 && -(a_4\#e_3+b_4\#e_1)\x\#E_g, \\
 \#H_g &=& -A\#b_s\x\#B_g -(Ab_4+2B)\#E_g \nonumber\\  &&-(\#e_3\#a_s+\#e_1\#b_s)\.\#E_g.\fa
Expressing the Gibbsian medium conditions in the form
\e \am \#D_g\\ \#H_g\a = \amm \=\A & \=\E'\\ \=\M{}^{-1} & \=\B\a\. \am \#B_g\\ \#E_g\a, \f
where the medium dyadics are understood in the Gibbsian sense, their expressions become
\ea \=\A &=&2B\=I_g -\#e_3\x(\#a_s\x\=I_g)-\#e_1\x(\#b_s\x\=I_g), \l{AA}\\
   \=\E' &=& -(a_4\#e_3+b_4\#e_1)\x\=I_g, \\
   \=\M{}^{-1} &=& -A\#b_s\x\=I_g, \\
   \=\B &=& -(Ab_4+2B)\=I_g -(\#e_3\#a_s+\#e_1\#b_s). \l{BB}\fa
One must notice that the dyadics $\=\E'$ and $\=\M{}^{-1}$ are antisymmetric and, consequently, do not have inverses. Thus, it is not possible to express the skewon-axion medium equations in the form \r{DB}. 

The most general skewon-axion medium can be defined in terms of three-dimensional Gibbsian vectors and dyadics in the form \cite{IB,IBDB}
\ea \#D_g &=& \tr\=A\ \#B_g -\=A\.\#B_g - \#c\x\#E_g, \l{D} \\
    \#H_g &=& -\#g\x\#B_g - \=A{}^T\.\#E_g - a\#E_g, \l{H} \fa
where $\=A$ is a dyadic, $\#c,\#g$ are two vectors and $a$ is a scalar. Together they make $9+3+3+1=16$ scalar parameters. Comparing with the representation \r{AA} - \r{BB}, we can identify the quantities in \r{D} and \r{H} as
\ea \=A &=& B\=I_g +\#a_s\#e_3 + \#b_s\#e_1, \l{A}\\
 \#c &=& b_4\#e_1+ a_4\#e_3, \\
 \#g &=& A\#b_s, \\
 a &=& B + Ab_4  , \l{a}\fa
which shows us that the number of free parameters of the medium defined by \r{B} must be less than 16. Expressing the dyadic $\=A$ as 
\e \=A= B\#e_2\#e_2 + (B\#e_3+\#a_s)\#e_3 + (B\#e_1+\#b_s)\#e_1, \f
its definition is seen to require $1+3+3=7$ parameters in a given basis $\#e_i$, while $\#c$ and $\#g$ require 2 and 1 parameters, respectively. Thus, the total number of free parameters in the medium under consideration appears to be 10. The pure axion (PEMC) medium \r{PEMC} special case is obtained for
\e \=A=\frac{M}{2}\=I_s,\ \ \ \#c=0,\ \ \ \#g=0,\ \ \ a=M/2. \f

\section{Reflection from SHDB Boundary}

Let us consider the basic problem, reflection of a plane wave from the planar interface of the skewon-axion half space $z<0$, in terms of Gibbsian quantities corresponding to time-harmonic fields.

\subsection{Dispersion Equation}

Denoting for simplicity
\e \#p=\#k/\o, \f
a plane wave with the dependence $\exp(-j\o\#p\.\#r)$ satisfies
\e \#p\x\#E_g = \#B_g,\ \ \ \ \#p\x\#H_g=-\#D_g. \l{pEB}\f
To obtain a condition for a plane wave in the skewon-axion medium,  substituting \r{D} and \r{H} and eliminating $\#B_g$ in \r{pEB} leaves us with an equation of the form \cite{IB}
\e \=D(\#p)\.\#E_g=0, \f
where we denote
\e \=D(\#p) = \#q(\#p)\x\=I_g,\f
\e \#q(\#p) = (\#g\.\#p-a)\#p -\#c + \#p\.\=A.\f
Because $\det\=D(\#p)=0$ for any $\#p$, there is no dispersion equation to restrict the choice of $\#p$. Thus, a plane wave with any vector $\#k$ is possible in the skewon-axion medium. One can further show that $\#q(\#p)=0$ for any $\#p$ exactly when the medium has no skewon component, i.e., when it is a pure axion (PEMC) medium with $\#c=\#g=0$ and $\=A=a\=I_s$. In the following we exlude the pure axion medium from the analysis.

\subsection{Field Conditions}

The field vectors $\#E_g$ and $\#B_g$ of the plane wave can be expressed as
\ea \#E_g &=& \#q(\#p) = (\#g\.\#p-a)\#p -\#c + \#p\.\=A, \l{Egg}\\
 \#B_g &=& \#p\x\#E_g = -\#p\x\#c -\#p\.\=A\x\#p. \l{Bgg}\fa
Inserting \r{A} - \r{a}, the field expressions for the special skewon-axion medium become
 \ea \#E_g &=& (\#b_s\.\#p-b_4)(A\#p+\#e_1) +(\#a_s\.\#p-a_4)\#e_3, \\
 \#B_g &=& (\#b_s\.\#p-b_4)\#p\x\#e_1 +(\#a_s\.\#p-a_4)\#p\x\#e_3. \fa
The form of the components 
 \ea \#E_g\.\#e_2 &=& A(\#b_s\.\#p-b_4)(\#p\.\#e_2), \\
 \#B_g\.\#e_3 &=& -(\#b_s\.\#p-b_4)(\#p\.\#e_2),\fa
implies that, for the general vector $\#p$, the fields satisfy the condition
\e A\#e_3\.\#B_g + \#e_2\.\#E_g=0, \f
which coincides with \r{3Bg} when \r{AT} is taken into account. If $\#p$ satisfies $\#p\.\#e_2=0$, we obtain both  $\#B_g\.\#e_3=0$ and $\#E_g\.\#e_2=0$, which are actually the same condition because they are connected by 
\e \#e_3\.(\#p\x\#E_g-\#B_g)= (\#e_1\.\#p)(\#e_2\.\#E_g) - \#e_3\.\#B_g=0. \f
 
Inserting \r{Egg} and \r{Bgg} in \r{D} and \r{H}, with \r{A} - \r{a} and after some algebraic steps, the following expansions for the other field components are obtained, 
\ea \#e_3\.\#D_g &=& -(\#p\.\#e_2)((\#a_s\.\#p-a_4)(\#b_s\.\#e_3)+\nonumber\\
&&+(\#b_s\.\#p-b_4)(Ab_4+2B+\#b_s\.\#e_1))\\ 
 \#e_2\.\#H_g &=& -A(\#p\.\#e_2)((\#a_s\.\#p-a_4)(\#b_s\.\#e_3)+\nonumber\\
 &&+(\#b_s\.\#p-b_4)(Ab_4+2B+\#b_s\.\#e_1)).\fa
Thus, in the general case, the fields satisfy
 \e A\#e_3\.\#D_g -\#e_2\.\#H_g =0, \f
which coincides with \r{3Dg} when \r{AT} is taken into account. Again, for $\#p\.\#e_2=0$ we obtain $\#e_3\.\#D_g=0$ and  $\#e_2\.\#H_g=0$, which are the same condition.
 
It now follows that, inserting \r{AT}, the sum of incident and reflected plane-wave fields at the boundary of the isotropic medium satisfy the conditions 
\ea T\o(\#B_g^i+\#B_g^r)\.\#e_3 &=& -k_o(\#E_g^i+\#E_g^r)\.\#e_2, \\
    T\o(\#D_g^i+\#D_g^r)\.\#e_3 &=& k_o(\#H_g^i+\#H_g^r)\.\#e_2, \fa 
which can be rewritten as
 \ea T(\#k^i\x\#E_g^i+\#k^r\x\#E_g^r)\.\#e_3 &=& -k_o(\#E_g^i+\#E_g^r)\.\#e_2, \\
    T(\#k^i\x\#H_g^i+\#k^r\x\#H_g^r)\.\#e_3 &=& -k_o(\#H_g^i+\#H_g^r)\.\#e_2. \fa 
Applying the relations
\e \#H_g^i = \frac{1}{\o\M_o}\#k^i\x\#E_g^i,\ \ \ \ \#H_g^r = \frac{1}{\o\M_o}\#k^r\x\#E_g^r, \f
\e \#k^r= (\#e_1\#e_1+\#e_2\#e_2-\#e_3\#e_3)\.\#k^i, \f
\e \#k^i\.\#E_g^i=0,\ \ \ \ \#k^r\.\#E_g^r=0, \f
\e \#k^r\.\#k^r = \#k^i\.\#k^i=k_o^2 , \f
one can find relations between the fields $\#E_g^i$ and $\#E_g^r$ which depend on the medium parameters and the vector $\#k^i$ of the incident wave. Skipping the algebraic details, we obtain the following result for the field components parallel to the boundary:
\e  \cal{A} \am E_1^r\\ E_2^r\a = \cal{B}\am E_1^i\\ E_2^i\a, \f
with matrices defined by
\e \cal{A}=\amm \A  & -\B \\ \g  & \de\a,\ \ \ \cal{B}=\amm -\A  & \B \\ \g  & \de\a, \f
\ea \A  &=& T k_2^i \\
    \B  &=& k_o + T k_1^i \\
    \g  &=& k_o^2 + Tk_ok_1^i - k_2^{i2} \\
    \de &=& (k_1^i + Tk_o)k_2^i. \fa
The determinants of the two matrices become
\ea \De &=&\det\cal{A}=-\det\cal{B} \nonumber\\ 
 &=& k_o((Tk_o+ k_1^i)^2 - (T^2-1)k_3^{i2}). \fa 
The reflected field components can be solved as
\ea E_1^r &=& -E_1^i +\frac{2\B}{\De}(\g E_1^i+ \de E_2^i) \\
    E_2^r &=&  - E_2^i+ \frac{2\A }{\De}(\g E_1^i + \de E_2^i) \\
    E_3^r &=& E_3^i +\frac{2}{k_3^i\De}(\A k_2^i+\B k_1^i)(\g E_1^i+\de E_2^i) \fa    

As a check, for $T=0$ we obtain  
\e E_2^r=-E_2^i,\ \ \ \ E_1^r= E_1^i + \frac{2k_1^ik_2^i}{k_1^{i2}+k_3^{i2}}E_2^i. \f
These coincide with the SH conditions because the latter relation equals $\#e_2\.\#H_g^r=-\#e_2\.\#H_g^i$, as can be verified. Also, for $T\ra\oo$ we obtain $E_3^r\ra-E_3^i$ and $H_3^r\ra-H_3^i$ which correspond to the respective DB conditions $\#e_3\.\#D_g\ra0$ and $\#e_3\.\#B_g\ra0$ for the total fields.

As another check we may consider the normally incident plane wave with $\#k^i=-k_o\#e_3$. In this case we obtain $E_1^r=E_1^i$ and $E_2^r=-E_2^i$, which coincide with the SH conditions. This is independent of any finite value of the parameter $T$. Thus, if $1/T$ is a small quantity, the SHDB boundary acts as a DB boundary, except for normal incidence where the SH condition suddenly takes over. 

Finally, for a plane wave satisfying $\#e_2\.\#k^i=0$ both SH and DB conditions are simultaneously satified for any $T$ and $k_1^i$ values.

\subsection{Eigenfields}

To find the eigenpolarizations satisfying
\e E_1^r=\la E_1^i,\ \ \ \ \ E_2^r=\la E_2^i, \f
we must solve 
\e \det({\cal{A}}-\la{\cal{B}})=0, \f
for $\la$. The solutions are simply
\e \la_\pm = \pm 1, \f
whence the eigenfields satisfy either PMC ($\la_+$) or PEC ($\la_-$) conditions at the SHDB boundary. The eigenfields satisfy
\ea \A_1E_{1+}^i &=& \B E_{2+}^i, \\
    \g E_{1-}^i &=& - \de E_{2-}^i. \fa
One can show that these equal the conditions
\ea T\o\#e_3\.\#B_{g+}^i + k_o\#e_2\.\#E_{g+}^i &=&0 \l{B+}\\
    T\o\#e_3\.\#D_{g-}^i -k_o\#e_2\.\#H_{g-}^i &=&0, \l{D-}\fa
which restrict the polarization of the two eigenwaves. Considering the two special cases, for $T=0$ (SH boundary) the eigenfields are either TE or TM with respect to $\#e_2$ while for $T\ra\oo$ (DB boundary) they are TE or TM with respect to $\#e_3$. Actually, in an isotropic medium, the polarization conditions are of the form
\ea \#a_+\.\#E_{g+} + \#b_+\.\#H_{g+} &=& 0\\
    \#a_-\.\#E_{g-} + \#b_-\.\#H_{g-} &=& 0, \fa 
which can be understood as generalizations of TE/TM conditions. Such polarization conditions have been previously encountered in media known by the name decomposable media, in which any fields can be decomposed in two non-interacting components \cite{deco}.

Since \r{B+} and \r{D-} are also valid for the respective reflected eigenwaves, they are valid for the total eigenfields. The conditions are linear in the field vectors and independent of the $\#k$ vector of the plane wave. Thus, they are valid for linear combinations of plane waves and, actually, for any fields. Thus, any field can be decomposed in two eigenfields satisfying either \r{B+} or \r{D-} for which the SHDB boundary can be replaced by respective PMC and PEC boundaries.

To show that any given field can be expanded as a sum of eigenfields, let us consider the general plane wave and express the conditions \r{B+} and \r{D-} as
\ea T\#e_3\.\#k\x\#E_{g+} + k_o\#e_2\.\#E_{g+} &=&0 \l{B1+}\\
    Tk_o\#e_3\.\#E_{g-} -\#e_2\.(\#k\x\#E_{g-}) &=&0, \l{D1-}\fa
which are of the form
 \e \#c_3\.\#E_{g+}=0,\ \ \ \ \ \#c_2\.\#E_{g-}=0, \l{cE}\f
with
 \ea \#c_3 &=& T\#e_3\x\#k + k_o\#e_2, \\
     \#c_2 &=& Tk_o\#e_3 -\#e_2\x\#k. \fa
These vectors satisfy
\e \#e_3\.\#c_3=0,\ \ \ \ \#e_2\.\#c_2=0. \f          
Expanding $\#a=\#k\x((\#c_3\x\#c_2)\x\#E_g)$ in two ways as
\ea \#a&=& \#k\x\#c_2(\#c_3\.\#E_g) -\#k\x\#c_3(\#c_2\.\#E_g)) \\
   &=& (\#c_3\x\#c_2)(\#k\.\#E_g) - (\#k\.(\#c_3\x\#c_2))\#E_g, \fa
and applying $\#k\.\#E_g=0$, the eigenfield representation
\e \#E_g = \#E_{g+} + \#E_{g-}, \f
is obtained by defining
\ea \#E_{g+} &=&  \frac{\#k\x\#c_3\#c_2}{\#k\.(\#c_2\x\#c_3)}\.\#E_g, \\
    \#E_{g-} &=&  \frac{\#k\x\#c_2\#c_3}{\#k\.(\#c_3\x\#c_2)}\.\#E_g. \fa
The expansion requires $\#k\.(\#c_3\x\#c_2)\not=0$ which corresponds to
\e T^2(k_1^2+k_2^2)+ k_1^2+k_3^2  \not=0. \f

For the magnetic field $\#H_g=\#H_{g+}+ \#H_{g-}$, the polarization conditions can be shown to take the form 
  \e \#c_2\.\#H_{g+}=0,\ \ \ \ \ \#c_3\.\#H_{g-}=0, \l{cH}\f
whence the magnetic eigenfields obey the expressions
\ea \#H_{g+} &=&  \frac{\#k\x\#c_2\#c_3}{\#k\.(\#c_3\x\#c_2)}\.\#H_g, \\
    \#H_{g-} &=&  \frac{\#k\x\#c_3\#c_2}{\#k\.(\#c_2\x\#c_3)}\.\#H_g. \fa
Obviously, the two eigenfields are dual to each other because the transformation $\#E_g\ra\#H_g,\ \#H_g\ra\#E_g$ corresponds to $\#E_{g+}\ra\#E_{g-},\ \#E_{g-}\ra\#E_{g+}$.

\begin{figure}[htb]
\centering
   \psfrag{x}[][]{${\#e}_1$} 
   \psfrag{y}[][]{${\#e}_2$} 
   \psfrag{z}[][]{${\#e}_3$} 
   \psfrag{t}[][]{$\theta$} 
   \psfrag{f}[][]{$\varphi$} 
   \psfrag{ki}[][]{${\#k}^i$} 
   \psfrag{kr}[][]{${\#k}^r$} 
\includegraphics[width=9cm]{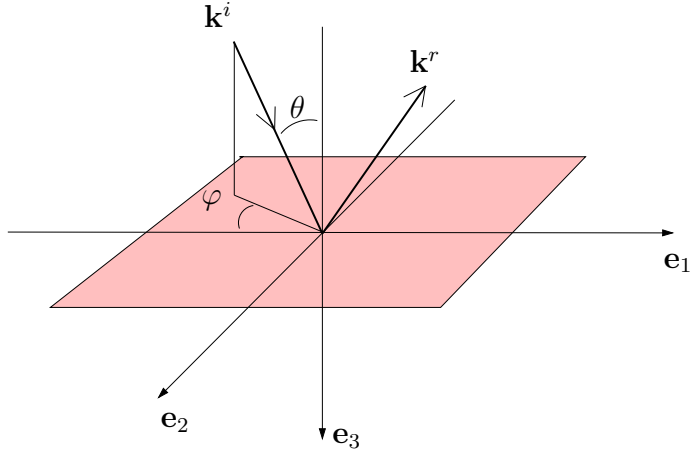}
\caption{\label{fig:geometry} Geometry of the reflection problem. A plane wave is incident on the SHDB boundary with directions defined by the angles $\theta,\varphi$ of the spherical coordinate system. The fields satisfy $\#k^i\.\#E_g^i=0$ and $\#k^r\.\#E_g^r=0$.}
\end{figure}

\section{Numerical example}

To have an idea of SHDB boundary in action, let us consider as an example a plane wave incident to the SHDB plane with $\VF=\pi/2$ in Figure \ref{fig:geometry}. Thus, the incident wave vector satisfies  
\e \#e_1\.\#k^i=0,\ \ \ \ \#k^i = k_o(\#e_2\sin\TH + \#e_3\cos\TH). \f
The relation between the incident and reflected field components transverse to $\#e_3$ can be represented by the planar reflection dyadic $\=R$ as
\e \#e_1E_1^r+ \#e_2E_2^r = \=R\.(\#e_1E_1^i+ \#e_2E_2^i), \f
\e \=R = \#e_1\#e_1 R_{11} + \#e_1\#e_2 R_{12} + \#e_2\#e_1R_{21} + \#e_2\#e_2R_{22}, \f
with
\ea R_{11} &=& \frac{\cos^2\TH - T^2\sin^2\TH}{\cos^2\TH + T^2\sin^2\TH},\\
 R_{12} &=& \frac{2T\sin\TH}{\cos^2\TH + T^2\sin^2\TH} , \\
 R_{21} &=& \frac{2T\sin\TH\cos^2\TH}{\cos^2\TH + T^2\sin^2\TH},\\
 R_{22} &=& -R_{11}. \fa
 The reflection dyadic satisfies
 \e \tr\=R=R_{11}+ R_{22}=0,\f
 \e \det\=R= -R_{11}^2-R_{12}R_{21}=-1, \f
 as is also obvious from the eigenvalues $\la_\pm=\pm1$.
 
\begin{figure}[t]
	{\includegraphics[width=9cm]{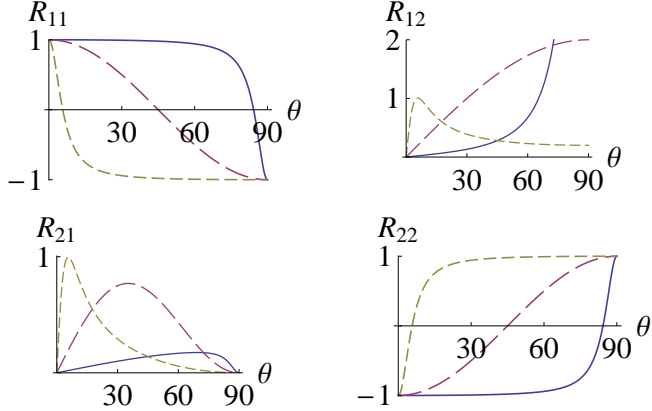}}
\caption{Components of the reflection dyadic $R_{ij}$ for a plane wave incident as $\VF=\pi/2$. The angle of incidence $\TH$ is shown in degrees. Curves in each figure correspond to parameter values $T=0.1$ (solid blue line); $T=1$ (long-dashed red line); and $T=10$ (short-dashed green line). Note that the co-polarized reflection coefficients satisfy $R_{11}=-R_{22}$.}
\label{fig:Rxy1a}
\end{figure}

Dependence of each component $R_{ij}$ of the reflection dyadic on the angle of incidence $\TH$ is depicted in Figure \ref{fig:Rxy1a} for three parameter values $T=(0.1, 1, 10)$. According to the previous theory, for $T=0.1$ the SHDB boundary should act almost like the DB boundary while for $T=10$ it should rather resemble the SH boundary. This is clearly seen from the graphs of $R_{11}$ and $R_{22}$ because for $T=0.1$ (solid curves) the component $E_2$ reflects almost from PEC with $R_{22}\approx -1$ and the component $E_1$ almost from PMC with $R_{11}\approx +1$. There is a deviation for angles close to grazing. The opposite is the case for $T=10$ with deviation close to normal incidence which is a known effect for the DB boundary. There are small cross-polarized reflections for small and large values of $T$. 

For $T=1$ the SHDB boundary exhbits properties not shared by SH or DB boundaries. In this case we have
\e R_{11}=-R_{22}=\cos2\TH,\ \ \ R_{12}=2\sin\TH,\ \ \ R_{21}=\sin2\TH\cos\TH. \f
For example, for the incidence angle $\TH=\pi/4$ the reflected field appears totally cross polarized, $R_{11}=R_{22}=0$. This is due to the nonreciprocal character of the SHDB boundary, which stems from the gyrotropic medium dyadics of the skewon-axion medium, \r{AA} -- \r{BB}.

As an another example we could consider the case $\#e_2\.\#k^i=0$, or $\VF=0$ in Figure \ref{fig:geometry}. Forming the reflection dyadic, we would obtain $R_{11}=-R_{22}=1$ and $R_{12}=R_{21}=0$ for all $\TH$ and $T$ values. This would make a dull diagram. As was already pointed out, the case $\#e_2\.\#k^i=0$ corresponds to reflection from either an SH boundary or a DB boundary, both of which yield the same result. Actually, the DB boundary can be characterized as an isotropic soft surface \cite{IBDB}, while the SH boundary is a soft surface for waves arriving from a certain direction, which in the present case corresponds to $\#e_2\.\#k^i=0$.

\section{Conclusion and Discussion}

The class of skewon-axion media can be described mathematically in natural (coordinate-independent) manner applying four-dimensional formalism. Instead of the 36 medium parameters of the most general linear medium, represented by a $6\x6$ matrix, the skewon-axion medium only involves 16 parameters defined by its characteristic $4\x4$ matrix. General solutions to the Maxwell equations can be expressed in simple form in the four-dimensional form. Considering a plane wave in terms of three-dimensional Gibbsian vector fields, it is shown that there is no characteristic equation for the vector $\#k$ in the medium which leaves a lot of freedom for the definition of the wave. 

In this paper a certain class of skewon-axion media is studied whose number of parameters is limited to 10. Such a medium has three specified axes denoted by the constant orthogonal unit vectors $\#e_1,\#e_2$ and $\#e_3$. It is shown that, in such a medium, two scalar conditions \r{3Bg}, \r{3Dg} are satisfied by the four Gibbsian field vectors. The form of the conditions suggests that, if there is a planar interface orthogonal to the vector $\#e_3$, \r{3Bg} and \r{3Dg} serve as boundary conditions for the fields. Since the conditions generalize those obtained for the soft-and-hard (SH) boundary and for the DB boundary for the limiting cases of the free parameter $T$, the novel boundary has been dubbed as the SHDB boundary.

Plane wave reflection from a SHDB boundary was considered and analytical expressions for the reflected field components were derived. It was shown that, for two eigenpolarizations, the SHDB boundary can be replaced by effective PEC and PMC boundaries just like for the SH and DB boundaries. The theory was tested by a numerical example demonstrating the nonreciprocal property of the SHDB boundary.

Although it appears quite easy to define the generalized boundary conditions \r{3Bg} and \r{3Dg}, quite simple in form, one may note that their realization in terms of a medium interface is not so obvious. In fact, it appears that, without the four-dimensional formalism, the introduction of the skewon-axion medium is a difficult task. Also, it is not at all obvious that a medium defined by \r{D} and \r{H} with \r{A} --\r{a} will yield a set of boundary conditions since it does not pop out directly from the medium equations but requires consideration of fields in the medium.

Realization of the skewon-axion medium by some metamaterial is a challenge for which there may not exist an easy solution. However, since the boundary conditions \r{3Bg} and \r{3Dg} appear more important than the medium itself, it should be easier to realize the SHDB boundary, whenever a useful engineering application for the boundary has been discovered. A corresponding realization for the DB boundary was recently found in terms of a planar structure containing metamaterial inclusions \cite{Zaluski}.

\section*{Appendix: Gibbsian Representation of Forms}

The connection between spatial multiforms and Gibbsian vectors can be expressed according to the following rules which are based on the spatial metric dyadic
\e \=I_g = \#e_1\#e_1 + \#e_2\#e_2+ \#e_3\#e_3, \f
which serves as the Gibbsian unit dyadic. Thus, a spatial one-form $\%\A_s=\A_1\ve_1 + \A_2 \ve_2+ \A_3\ve_3$ is transformed to a Gibbsian vector as
\e \%\A_s \ \ \ra\ \ \%\A_g = \=I_g|\%\A_s = \A_1\#e_1 + \A_2\#e_2+ \A_3\#e_3. \f
Similarly, a spatial two-form $\#B = B_{12}\ve_{12}+ B_{23}\ve_{23} + B_{31}\ve_{31}$ is transformed as
\e \#B\ \ \ra\ \ \#B_g = \#e_{123}\L\#B = B_{12}\#e_3+ B_{23}\#e_1 + B_{31}\#e_2. \f
The bar product of a spatial vector $\#a_s$ and a spatial one-form $\%\A_s$ transforms to a dot-product of Gibbsian vectors,
\e \#a_s|\%\A_s \ \ \ra\ \ \#a_s\.\%\A_g = \#a_s\.\=I_g|\%\A_s = \#a_s|\%\A_s. \f
The wedge product of two spatial one-forms $\%\A_s,\%\B_s$ becomes the cross product of two Gibbsian vectors,
\e \%\A_s\W\%\B_s\ \ \ra\ \ \%\A_g\x\%\B_g = \#e_{123}\L(\%\A_s\W\%\B_g), \f
and the wedge product of three one-forms transforms to a scalar,
\e \%\A_s\W\%\B_s\W\%\g_s\ \ \ra\ \ \%\A_g\.(\%\B_g\x\%\g_g)=  \%\A_s|(\#e_{123}\L(\%\B_s\W\%\g_s)). \f
Finally, the contraction product of a spatial vector $\#a_s$ and a spatial two-form $\#B$ is transformed to the cross product of the Gibbsian vectors $\#a_s$ and $\#B_g$ as
\e \#a_s\J\#B\ \ra\ \#a_s\x\#B_g = \=I_g|(\#a_s\J\#B). \f


\begin{thebibliography}{99}


\bibitem{Deschamps} G.A. Deschamps, "Electromagnetics and differential forms," {\it Proc.\ IEEE}, vol.69, no.6, pp.676--696, 1981.

\bibitem{Difform} I.V. Lindell, {\it Differential Forms in Electromagnetics,} New York: Wiley, 2004. 

\bibitem{Hehl} F.W. Hehl and Yu.\ Obukhov, {\it Foundations of Classical Electrodynamics}, Boston: Birkh\"auser, 2004.

\bibitem{PEMC} Lindell, I.~V., and A.~Sihvola, "Perfect electromagnetic conductor", {\it J. Electro.\ Waves Appl.} Vol.~19, No.~7, 861--869, 2005.

\bibitem{PEMC1} I.V. Lindell, A. Sihvola, ``Transformation method for problems involving perfect electromagnetic (PEMC) structures,'' {\it IEEE Trans.\ Antennas Propag.}, vol.53, no.9, pp.3005--3011, September 2005.

\bibitem{AnnPhys} A. Sihvola and I.V. Lindell, "Perfect electromagnetic conductor medium", {\it Ann.\ Phys.\ (Berlin)}, vol.17, no.9-10, pp.787--802, 2008.

\bibitem{IB} I.V. Lindell, "The class of bi-anisotropic IB-media," {\it Prog.\ Electromag.\ Res.}, vol.57, pp.1--18, 2006. 

\bibitem{Hehl02} Hehl,~F.~W., Yu.N. Obukhov and G.F. Rubilar, ``On a possible new type of a T odd skewon field linked to electromagnetism," {\it ArXiv\/} gr-qc/020315 (14 pages) 2002.

\bibitem{Obukhov04} Obukhov,~Yu.~N.~and F.W. Hehl, ``On possible skewon effects on light propagation," {\it Phys.\ Rev.\ D\/}, Vol.~70, 125015, 2004.

\bibitem{Hehl05} Hehl,~F.~W., Yu.~N.~Obukhov, G.~F.~Rubilar and M.~ Blagojevic, ``On the theory of the skewon from electrodynamics to gravity," {\it Phys.\ Lett.\ A\/} Vol.~ 347, pp.14--24, 2005.

\bibitem{IBDB} I.V. Lindell, A.H. Sihvola, ``Uniaxial IB-medium interface and novel boundary conditions,'' {\it IEEE Trans.\ Antennas Propagat.}, vol.57, no.3, pp.694--700, March 2009.

\bibitem{DB} I.V. Lindell and A. Sihvola: "Electromagnetic boundary condition and its realization with anisotropic metamaterial," {\it Phys.\ Rev.\ E}, vol.79, no.2, 026604 (7 pages), 2009.

\bibitem{AP10} I.V. Lindell and A. Sihvola, "Electromagnetic boundary conditions defined in terms of normal field components," {\it Trans.\ IEEE Antennas Propag.}, vol.58, no.4, pp.1128--1135, April 2010.

\bibitem{Rumsey} V.H. Rumsey, "Some new forms of Huygens' principle," {\it IRE Trans.\ Antennas Propagat.}, vol.7, Special supplement, pp.S103--S116, 1959.

\bibitem{Yee} K.S. Yee, "Uniqueness theorems for an exterior electromagnetic field," {\it SIAM J.\ Appl.\ Math.}, vol.18, no.1, pp.77--83, 1970.

\bibitem{Kress} R. Kress, "On an exterior boundary-value problem for the time-harmonic Maxwell equations with boundary conditions for the normal components of the electric and magnetic field," {\it Math.\ Meth.\ in the Appl.\ Sci.}, vol.8, pp.77--92, 1986.

\bibitem{Kong08} B. Zhang, H. Chen, B.-I. Wu, J.A. Kong, "Extraordinary surface voltage effect in the invisibility cloak with an active device inside," {\it Phys.\ Rev.\ Lett.}, vol.100, 063904 (4 pages), February 15, 2008.

\bibitem{Yaghjian}A.D. Yaghjian and S. Maci, "Alternative derivation of electromagnetic cloaks and concentrators," {New J. Phys.}, vol.10, 115022 (29 pages), 2008. Corrigendum, {\it ibid}, vol.11, 039802 (1 page), 2009.

\bibitem{Weder} R. Weder, "The boundary conditions for point transformed electromagnetic invisible cloaks," {\it J. Phys.\ A}, vol.41, 415401 (17 pages), 2008.

\bibitem{Kildal09} P.-S. Kildal, ``Fundamental properties of canonical soft and hard surfaces, perfect magnetic conductors and the newly introduced DB surface and their relation to different practical applications included cloaking,'' {\it Proc.\ ICEAA'09}, Torino, Italy Aug. 2009, pp.607--610.

\bibitem{AP09} I.V. Lindell, A. Sihvola, P. Yl\"a-Oijala and H. Wall\'en, "Zero backscattering from self-dual objects of finite size," {\it IEEE Trans.\  Antennas Propag.}, vol.57, no.9, pp.2725--2731, September 2009.

\bibitem{SS} I.V. Lindell and A. Sihvola, "Simple skewon medium realization of DB boundary conditions", {\it Prog.\ Electro.\ Res.}, submitted.

\bibitem{SHS1} P.-S. Kildal, "Definition of artificially soft and hard surfaces for electromagnetic waves," {\it Electron.\ Lett.}, vol.24, pp.168--170, 1988.

\bibitem{SHS2} P.-S. Kildal and A. Kishk, "EM modeling of surfaces with stop or go characteristics - artificial magnetic conductors and soft and hard surfaces,'' {\it ACES Journal}, vol.18, no.1, pp.32--40, 2003.

\bibitem{deco} I.V. Lindell and F. Olyslager, "Generalized decomposition of electromagnetic fields in bi-anisotropic media," {\it IEEE Trans.\ Antennas Propag.}, vol.46, no.10, pp.1584--1585, October 1998.

\bibitem{Zaluski} D. Zaluski, D. Muha and S. Hrabar, "DB boundary based on resonant metamaterial inclusions," {\it Metamaterials'2011}, Barcelona, October 2011, pp.820--822.


\end{thebibliography}
\end{document}